\documentclass[article,twocolumn]{revtex4}

\usepackage{multirow} % necesario para fusionar filas
\usepackage{booktabs}
\usepackage{enumitem}
\usepackage{amsmath}   
\usepackage{xcolor}
\usepackage[utf8]{inputenc}
\usepackage{empheq}
\usepackage{graphicx}
\usepackage{subcaption}
\usepackage{physics}
\usepackage[a4paper]{geometry}
\usepackage{setspace}
\usepackage{ragged2e} % <- Necesario para \justifying
\usepackage{caption}
\DeclareCaptionJustification{justified}{\justifying}
\begin{document}

\setstretch{1.08} 
\raggedbottom

\title{Implementing a Universal Set of Geometric Quantum Gates through Dressed-State assisted STA}
\author{M. Estefanía Rus$^1$, Alejandro Ferrón$^1$, Omar Osenda$^2$
and Sergio S. Gomez$^1$}

\address{$^1$Instituto de Modelado e Innovación Tecnológica, Universidad Nacional del Nordeste, CONICET, Facultad de Ciencias Exactas y Naturales y Agrimensura, Avenida Libertad 5400, W3404AAS Corrientes, Argentina.}
\address{$^2$Instituto de Física Enrique Gaviola, Universidad Nacional de Córdoba, CONICET, Facultad de Matemática, Astronomía, Física y Computación, Av. Medina Allende s/n, Ciudad Universitaria, CP:X5000HUA Córdoba, Argentina }

\date{\today}

\begin{abstract}
Geometric quantum computation relies on the geometric phase that arises in adiabatic cyclic evolutions of non-degenerate quantum systems, enabling the design of robust quantum gates. However, the adiabatic condition requires long evolution times, making the system vulnerable to decoherence. In this work, we propose a scheme to realize fast and high-fidelity geometric quantum gates by applying the Superadiabatic Transitionless Driving (SATD) protocol within the dressed-state framework. We analyze the implementation of single-qubit gates in a two-level system driven by a microwave field, focusing in particular on the NV center in diamond. We show how the dynamical phase can be canceled to obtain purely geometric operations. The robustness of the gates is assessed under systematic errors and environmental decoherence, demonstrating high fidelities even in regimes with strong fluctuations. Finally, we extend the protocol to construct nontrivial two-qubit gates, highlighting its feasibility for scalable quantum information processing.

\end{abstract}

\maketitle

\section{Introduction}

Quantum gates are the fundamental building blocks of quantum computation, enabling the manipulation of qubit states through controlled unitary operations \cite{Nielsen}. 
In recent years, geometric quantum gates (GQG) have attracted significant attention due to their intrinsic robustness against certain types of local noise and control errors. These gates are based on the adiabatic theorem, which ensures that a quantum system evolving slowly enough remains in its instantaneous eigenstate. In this context, the system acquires a phase composed of two parts: a dynamical phase that depends on the energy of the state and the duration of the evolution, and a geometric phase that depends only on the trajectory followed in parameter space. The latter, known as Berry’s phase \cite{Berry1984}, has been extensively studied as a resource for fault-tolerant quantum information processing \cite{Zanardi1999, Ekert(2000), GQG-Review2023, Zhu2002, Wu2013}.

However, despite its inherent robustness against control errors \cite{Lupo2007, Chen2020}, geometric quantum computation typically relies on adiabatic evolution, which imposes a major limitation: the need for slow dynamics increases the total gate time, which in turn heightens the effects of decoherence and reduces fidelity. These drawbacks have been addressed through the development of shortcuts to adiabaticity (STA) strategies \cite{STAreview(2019), BerryTQD, LR-Invariants(2012)}, such as Tansitionless Quantum Driving (TQD) \cite{PRA93(2016)} and Invariant-based reverse engineering \cite{NoncyclicGG(2021)}, which enable faster gate implementations while retaining the geometric nature and robustness of the operations. Another such method is the Superadiabatic Transitionless Driving (SATD) technique, introduced by Basik et al. in 2016 \cite{Baksic(2016)}. This approach, based on dressed states, generalizes the standard TQD and has been primarily applied to STIRAP-like systems \cite{Coto-SATD, Zhou-SATD, Ribeiro-Clerk(2019), Liu(2017), Zhang2020}, where a suitable choice of the dressed-state basis helps to suppress the population of the intermediate state. 

GQG have been implemented in a wide range of platforms, including superconducting circuits \cite{Li2021, Yang2023, Wang2018, Liang2024}, semiconductor quantum-dots \cite{Ma2024}, Rydberg atoms \cite{Zhao2017, Jin2024} and nitrogen-vacancy (NV) centers in diamond \cite{Cheng2021, Chen2025-arxiv}. NV centers, in particular, have attracted considerable attention due to their unique properties: they exhibit long-lived spin coherence, optical addressability, and operation under ambient conditions \cite{Doherty2013}. Both dynamical and geometric quantum gates have been experimentally demonstrated using NV centers, showing that they are highly suitable for a variety of quantum applications \cite{Arroyo2014, Sekiguchi2017, Bartling2025, Kleißler2018, Ma2023, Zu2014}. 

In this paper, we propose the implementation of fast GQG by applying the SATD strategy to a two-level system. 
We consider the NV center in its ground state as a physical platform for the qubit, which, in the presence of an external static magnetic field, allows the spin-1 system to be effectively reduced to a two-level subspace composed of its lowest energy states. The system is driven by a microwave field, with pulses designed such that the Hamiltonian drives the state along an orange-slice path (OSP) on Bloch's sphere, ensuring the accumulation of a purely geometric phase.
The SATD protocol is then applied to speed up the evolution, but we find that the corrected pulses generated by this approach induce an undesired dynamical phase. However, we show that by exploiting the freedom in choice of the dressed-state basis, this phase can be eliminated, allowing for the implementation of the desired geometric gates.
We study the influence of decoherence, due to the interaction of the electron spin with its environment, and of systematic errors produced by imperfections in the control pulses, showing that the approach remains robust and suitable for reliable quantum gate implementation. Furthermore, we extend the protocol to design a nontrivial two-qubit gate based on the same STA strategy developed for single-qubit operations, demonstrating the viability of this method for practical quantum computing applications.

\section{Model}

We consider a two-level system defined in the basis of states $|0\rangle$ and $|1\rangle$ with an energy difference of $\hbar\omega_e$,  which are coupled via a microwave (MW) field of frequency $\omega(t)$. In the rotating wave approximation (RWA), the hamiltonian is given by
\begin{equation}
    H_0 (t)=\frac{\hbar}{2} \begin{pmatrix}
        \Delta(t) && \Omega_R(t)e^{-i\varphi(t)} \\
        \Omega_R(t)e^{i\varphi(t)} && -\Delta(t)
    \end{pmatrix}
    \label{eq:H0}
\end{equation}
where $\Omega_R(t)$ is the Rabi frequency of the MW and $\Delta(t)=\omega_e-\omega(t)$ is the detuning. Then, the adiabatic eigenstates are 
\begin{subequations}
    \begin{equation}
        |\psi_+(t)\rangle = \cos{\frac{\theta(t)}{2}} |0\rangle + \sin{\frac{\theta(t)}{2}} e^{i\varphi(t)} |1\rangle
    \end{equation}
    \begin{equation}
        |\psi_-(t)\rangle = \sin{\frac{\theta(t)}{2}} e^{-i\varphi(t)} |0\rangle - \cos{\frac{\theta(t)}{2}} |1\rangle  
    \end{equation}
\label{eq:states}
\end{subequations}
with eigenvalues $E_{\pm}=\pm\Omega/2$, where $\Omega=\sqrt{\Omega_R^2+\Delta^2}$, and the time-dependent functions $\theta=\arctan{(\Omega_R/\Delta)}$ and $\varphi$ represent the polar and azimuthal angles that define the trajectory of the state in Bloch's sphere. 

Although the Hamiltonian from Eq.~\eqref{eq:H0} can describe the dynamics of a wide variety of systems, we consider here the NV center in diamond as the qubit platform. This system is inherently spin-1, but it can be effectively reduced to a two-level subspace under a sufficiently strong static magnetic field $B_z$, as it is shown in Fig.~1(a). 

%In this regime, the time-dependent pulses and their phases can be modeled as the effect of a microwave (MW) or radio-frequency (RF) field acting on the electronic spin.

If the system is initially prepared in the state $\ket{\psi_i}=a_+\ket{\psi_+(0)}+a_-\ket{\psi_-(0)}$, with $a_\pm=\braket{\psi_\pm(0)}{\psi_i}$, and undergoes a cyclic evolution over a time interval $T$, the dynamical phase is canceled. Then, the eigenstates acquire purely geometric phases, such that $\ket{\psi_\pm(T)}=e^{\pm i\gamma_g}\ket{\psi_\pm(0)}$. The resulting unitary transformation on the qubit is
\small
\begin{equation}
U(\chi,\gamma_g)=\begin{pmatrix}
        \cos{\gamma_g}+i \cos{\chi}\sin{\gamma_g} & i\sin{\chi}\sin{\gamma_g} \\
        i\sin{\chi}\sin{\gamma_g} & \cos{\gamma_g}-i\cos{\chi}\sin{\gamma_g} \end{pmatrix}
\label{eq:Ugeom}
\end{equation}
\normalsize
which represents a general geometric single-qubit gate that depends on the geometric phase $\gamma_g$ and $\chi=\theta(0)$. In this framework, the operator 
$U(\chi,\gamma_g)$ allows realizing specific gates: for $\chi=0$, one obtains a $z$-rotation, 
$U_z(\gamma_g) = \exp(i \gamma_g \sigma_z)$, and for $\chi = \pi/2$, a rotation around the $x$-axis, 
$U_x(\gamma_g) = \exp(i \gamma_g \sigma_x)$. These two gates form a universal set for single-qubit operations, enabling the construction of arbitrary quantum state transformations.

\begin{figure}[t!]
    \centering   \includegraphics[width=1\linewidth]{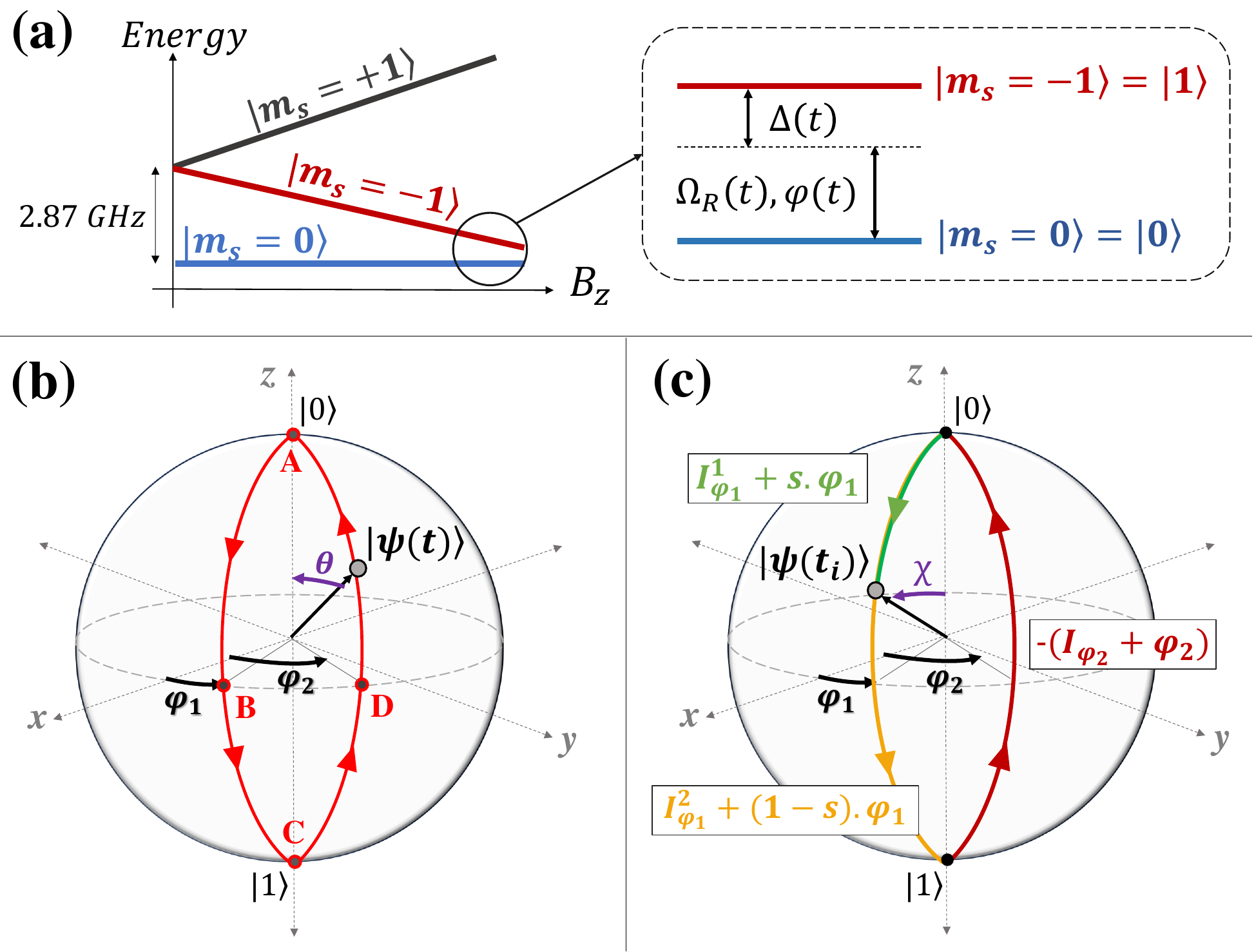}
\caption{(a) Energy levels of the NV spin ground state in the presence of a static magnetic field $B_z$. The inset shows the two lowest Zeeman levels that are taken as the qubit states $\ket{0}$ and $\ket{1}$. (b) Bloch's sphere showing the OS path followed by the state $\ket{\psi(t)}$. (c) Phases accumulated during the evolution with SATD in the DS frame. The system starts at $\theta(t_i) = \chi$ and $\varphi = \varphi_1$. It first evolves from this point to $\theta = \pi$ along the yellow path, acquiring a phase $I_{\varphi_1}^1 + s\varphi_1$, with $s=(1+\cos\chi)/2$. Then, it moves from $\theta = \pi$ to $\theta = 0$ with azimuthal angle $\varphi_2$ (red line), gaining a phase $I_{\varphi_2} + \varphi_2$. Finally, it returns from $\theta = 0$ to $\theta = \chi$ along the green path with azimuthal angle $\varphi_1$, acquiring a phase $I_{\varphi_1}^2 + (1-s)\varphi_1$. Here, $I_{\varphi_1}^{1,2}$ and $I_{\varphi_2}$ are the integrals of the DS energy along each segment of the trajectory.} 
    \label{fig:BlocSphere}
\end{figure}

In the case of $U_z(\gamma_g)$, we find the system initially at $\ket{\psi_+(0)}=\ket{0}$ and we make it evolve following the path ABCDA on Bloch's sphere, as depicted in Fig.~1(b), if
the time-dependent pulses $\Delta(t)$ and $\Omega_R(t)$  are defined as
\begin{subequations}
 \begin{equation} \label{eq:Delta(t)}
  \Delta(t)=  \left \{
      \begin{array}{cc}
          \Delta_0 \left[\cos{\left(\frac{\pi t}{\tau}\right)}+1\right] &  0\leq t<\tau \\[8pt]
          \Delta_0 \left[\cos{\left(\frac{\pi (t-\tau)}{\tau}\right)}-1\right] &  \tau\leq t < 2\tau \\[8pt] 
         \Delta_0 \left[\cos{\left(\frac{\pi (t-2\tau)}{\tau}\right)}+1\right] &  2\tau\leq t<3\tau \\[8pt]
         \Delta_0 \left[\cos{\left(\frac{\pi (t-3\tau)}{\tau}\right)}-1\right] &  3\tau\leq t\leq 4\tau \\
      \end{array}
   \right .
\end{equation}
%\vspace{-0.1cm}
\begin{equation} \label{eq:OmegaR(t)}
  \Omega_R(t)=  \left \{
      \begin{array}{rcl}
          \Omega_0 \left[1-\cos{\left(\frac{\pi t}{\tau}\right)}\right] &  0\leq t<\tau \\[8pt]
          \Omega_0 \left[1+\cos{\left(\frac{\pi (t-\tau)}{\tau}\right)}\right] &  \tau\leq t < 2\tau \\[8pt]
         \Omega_0 \left[1-\cos{\left(\frac{\pi (t-2\tau)}{\tau}\right)}\right] &  2\tau\leq t<3\tau \\[8pt]
         \Omega_0 \left[1+\cos{\left(\frac{\pi (t-3\tau)}{\tau}\right)}\right] &  3\tau\leq t\leq 4\tau 
      \end{array}
   \right .
  \end{equation}
\end{subequations}
For $U_x(\gamma_g)$, on the other hand, the initial state is $\ket{\psi_+(0)}=(\ket{0}+\ket{1})/\sqrt{2}$ and the system follows the path BCDAB. In this case, $\Delta(t)$ and $\Omega_R(t)$ are give by
\begin{subequations}
 \begin{equation}
  \Delta(t)=  \left \{
      \begin{array}{cc}
          \Delta_0 \left[\cos{\left(\frac{\pi t}{\tau}\right)}-1\right] &  0\leq t<\tau \\[8pt]
          \Delta_0 \left[\cos{\left(\frac{\pi (t-\tau)}{\tau}\right)}+1\right] &  \tau\leq t < 2\tau \\[8pt] 
         \Delta_0 \left[\cos{\left(\frac{\pi (t-2\tau)}{\tau}\right)}-1\right] &  2\tau\leq t<3\tau \\[8pt]
         \Delta_0 \left[\cos{\left(\frac{\pi (t-3\tau)}{\tau}\right)}+1\right] &  3\tau\leq t\leq 4\tau \\
      \end{array}
   \right .
\end{equation}
%\vspace{-0.1cm}
\begin{equation}
  \Omega_R(t)=  \left \{
      \begin{array}{rcl}
          \Omega_0 \left[1+\cos{\left(\frac{\pi t}{\tau}\right)}\right] &  0\leq t<\tau \\[8pt]
          \Omega_0 \left[1-\cos{\left(\frac{\pi (t-\tau)}{\tau}\right)}\right] &  \tau\leq t < 2\tau \\[8pt]
         \Omega_0 \left[1+\cos{\left(\frac{\pi (t-2\tau)}{\tau}\right)}\right] &  2\tau\leq t<3\tau \\[8pt]
         \Omega_0 \left[1-\cos{\left(\frac{\pi (t-3\tau)}{\tau}\right)}\right] &  3\tau\leq t\leq 4\tau 
      \end{array}
   \right .
  \end{equation}
\end{subequations}
In the construction of both gates, each time interval corresponds to a quarter of the OS path, and the phase $\varphi$ has a constant value through each interval.

%\vspace{-0.25cm}
\subsection{STA protocol with dressed-states} %Dressed-states fomralism}

The dressed-states approach introduced by Baksic $\textit{et al.}$ in \cite{Baksic(2016)} generalizes the counterdiabatic (CD) driving method by applying two successive unitary transformations to the system. The goal is to design a total Hamiltonian $H_{SATD}(t)=H_0(t)+H_c(t)$, such that the system evolves from an initial adiabatic state to a target adiabatic state, even when the evolution is fast.

The method starts in the Schrödinger picture where the system is governed by $H_0(t)$, and a control Hamiltonian $H_c(t)$ will be added later to ensure the desired evolution. The first step is to move to the adiabatic frame by applying the unitary transformation $U_{ad}(t)=\sum_{n}\ket{\psi_n(t)}\bra{\psi_n(0)}$, where $\ket{\psi_n}$ are given by Eq.~\eqref{eq:states}. The hamiltonian in this frame becomes
\begin{equation}
H_{ad}(t) = U_{ad}^\dagger(t)[H_0(t) + H_c(t)]U_{ad}(t) + i\hbar \dot{U}^\dagger_{ad}(t)U_{ad}(t)
  \label{eq:Had}
\end{equation}
with eigenstates $\ket{\psi_{ad,n}(t)}=U^\dagger_{ad}(t)\ket{\psi_n(t)}$. A second transformation is then applied with an operator $V(t)=\sum_n|\tilde{\psi}_n(t)\rangle \bra{\psi_n(0)}$, where $\{|\tilde{\psi}_n(t)\rangle \}$ are the so called dressed-states (DS). The resulting hamiltonian is
\begin{equation}
    H_{DS}(t)= V^\dagger H_{ad}V + V^\dagger U_{ad}^\dagger H_c U_{ad} V + i \hbar\dot{V}^\dagger V
    \label{eq:HDS}
\end{equation}
where the time dependence of all the operators has been omitted for simplicity.
The key idea of the protocol is to choose the basis $\{|\tilde{\psi}_n(t)\rangle \}$ such that the following conditions are satisfied:
\begin{itemize}[leftmargin=*]
\item[1.] $V(0)=V(T)=1$ so the DS coincide with the adiabatic states \eqref{eq:states} at the initial and final time of the evolution, up to a phase.
\item[2.] $H_{DS}(t)$ must be diagonal in the time-independent basis $\{ \ket{\psi_n(0)}\}$, so the dynamics in the DS frame becomes trivial. Transforming back to the Schrödinger picture, this corresponds to an evolution along the vectors $U_{ad}V|\psi_n(0)\rangle= U_{ad}|\tilde{\psi}_n(t)\rangle$.
\end{itemize}
This additional freedom allows us to engineer a feasible correction $H_c(t)$, such that undesirable transitions are avoided. Notice here that choosing $V(t)=1 \ \forall t$ reduces the method to the standard TQD.

Following the proposal presented in \cite{Baksic(2016)}, we define the DS unitary transformation operator as
\begin{equation}
    V(t)=\exp{-i\mu ~S_x } 
    \label{eq:V}
\end{equation}
and the corrective hamiltonian in the adiabatic basis as
\begin{equation}
    U_{ad}^\dagger H_c U_{ad}= g_x  S_x + g_z  S_z
    \label{eq:Hc}
\end{equation}
where $S_i$ ($i=x,y,z$) are the spin-1/2 operators and $\mu, \ g_x$ and $g_z$ are time-dependent functions defined to fulfill the previous conditions. By substituting Eqs.~\eqref{eq:V} and \eqref{eq:Hc} in Eq.~\eqref{eq:HDS}, and imposing the latter to be diagonal, we obtain the following relations
\begin{subequations}
    \begin{equation}
        g_x=-\dot{\mu} +\dot{\theta}\sin{\varphi}
    \end{equation}
    \vspace{-0.9cm}
    \begin{equation}
        g_z=-\Omega+ \frac{\dot{\theta}\cos{\varphi}}{\tan{\mu}}
    \end{equation}
\label{eq:gxgz}
\end{subequations}
We can determine the value of $\mu$ from the second equation, and $g_z$ is then an arbitrary function that can be designed to meet the specific demands of the desired evolution.

Finally, turning back to the original frame we get
\begin{equation} \label{eq:H_SATD}
\begin{split}
    H_{SATD}(t)&= H_0(t)+U_{ad}^\dagger(t)~H_c(t)~U_{ad}(t)\\
    &=\frac{\hbar}{2}\begin{pmatrix}
        \Tilde{\Delta}(t) & \Tilde{\Omega}_R (t) e^{-i\Tilde{\varphi}(t)} \\
        \Tilde{\Omega}_R (t) e^{i\Tilde{\varphi}(t)} & -\Tilde{\Delta}(t)
    \end{pmatrix}
\end{split}
\end{equation}
with
\begin{subequations} \label{eq:PulsesSTA}
    \begin{equation}
        \Tilde{\Delta}=(g_z+\Omega)\cos{\theta}+g_x \sin{\theta}\cos{\varphi}
    \end{equation}
    \vspace{-0.5cm}
    \begin{equation}
         \Tilde{\Omega}_R = \sqrt{ [(g_z+\Omega)\sin{\theta}-g_x\cos{\theta}\cos{\varphi}]^2 +  g_x^2 \sin^2{\varphi} }
    \end{equation}
    \vspace{-0.5cm}
    \begin{equation}
         \Tilde{\varphi} = \varphi + \arctan{\left( \frac{g_x \sin{\varphi}}{(g_z+\Omega)\sin{\theta}-g_x\cos{\theta}\cos{\varphi}} \right)}
    \end{equation}
\label{eq:ModPulses}
\end{subequations}
\noindent
The Hamiltonian \eqref{eq:H_SATD} has the same form as the original one, but with the new pulses given by Eqs.~(12). 

It should be noted that the original phase $\varphi$ is a constant that changes abruptly at $t=T/2$. Since the modified pulses depend on this phase, such a discontinuity can lead to unwanted effects in the dynamics. A convenient way to overcome this issue is to introduce a time-dependent phase that smoothly interpolates the abrupt jump, thereby ensuring continuity in the pulse profiles. In Appendix A, we show that the effects of this modulation do not affect significantly the fidelity of the protocol when choosing the appropriate constant parameters of the system.

On the other hand, it is well known that the pulses modified by STA techniques typically require larger amplitudes than the original ones, which is undesirable as this may introduce significant challenges for experimental implementation. However, this drawback can be mitigated by a suitable choice of the parameters. For this reason, in Fig.~2 we analyze the relation between the maximum amplitudes of the original pulses $\Delta$ and $\Omega_R$, and those of the SATD-modified $\tilde{\Delta}$ and $\tilde{\Omega}_R$. We define the parameters $\eta=\Delta_0/\Omega_0$ and $x=\tau \Omega_0$, and plot as a function of both the quotient $R=\max[z]/\max[\tilde{z}]$, with $z$ being $\Omega_R$ (panel a) and $\Delta$ (panel b).
In the following section we show that $R$ scales with $x$, which means that the operation time $\tau$ can be reduced without increasing $R$ by appropriately tuning $\Omega_0$. Consequently, by selecting suitable values of $x$ and $\eta$, one can design STA-corrected pulses that preserve low amplitudes, approaching the ideal case where $R \le 1$.

%\textcolor{red}{We show bellow in the next section that R scales with $x$, meaning that in order to decrease the $\tau$ without increasing $R$, we could increase $\Omega_0$. Besides that scaling, by setting a desired value of $x$ and $\eta$ we can also adjust the pulse amplitude to perform on a desired time $\tau=x/\Omega_0$. The results of the figure shows that one can always choose values of $\eta$ and $x$ with corrections to the pulse with low amplitudes which represents the ideal case , where the value of $R$ never surpasses the original one, namely $R\simeq 1$ }. 

\begin{figure}[t!]
    \centering
    \includegraphics[width=0.8\linewidth]{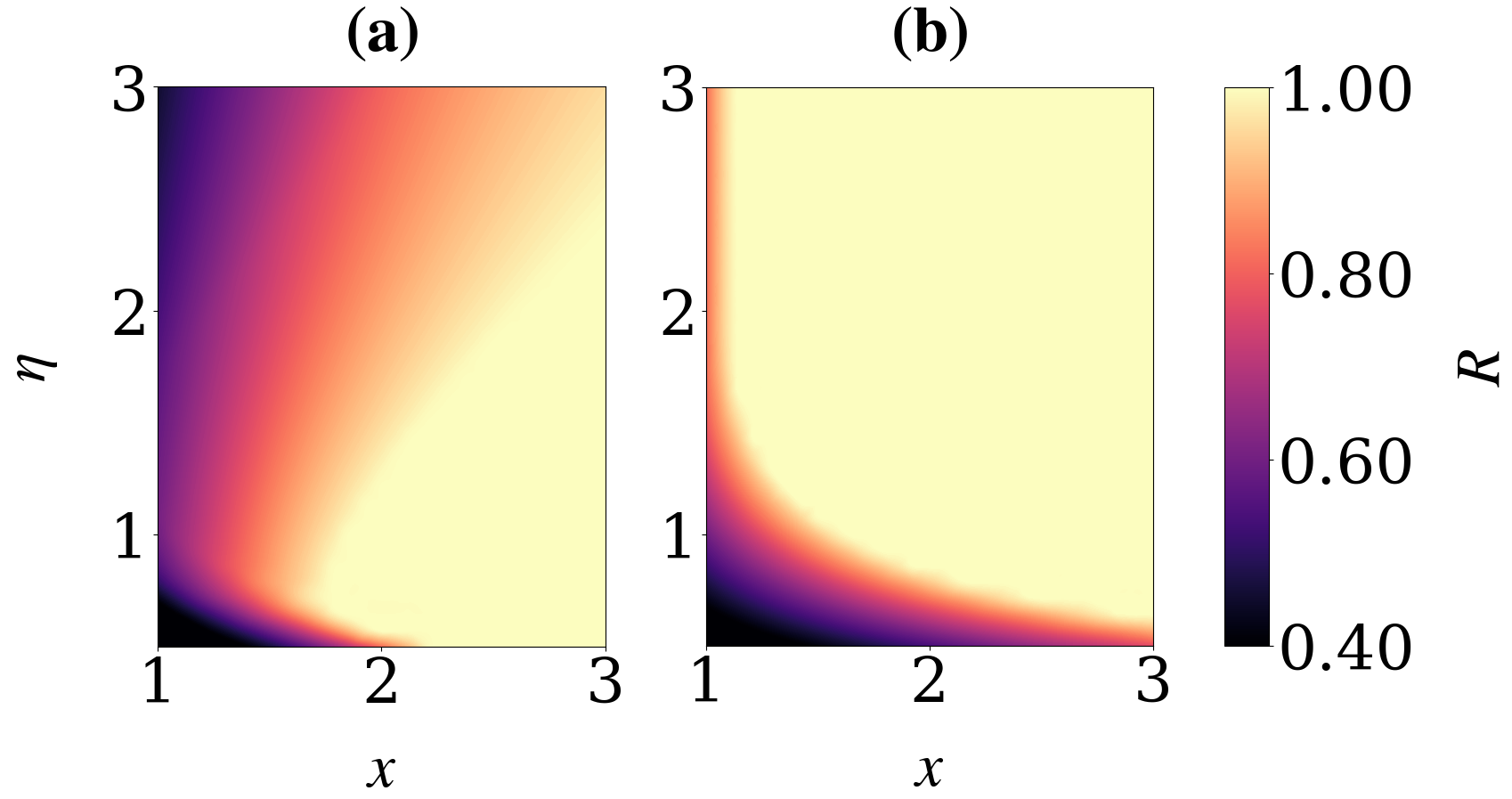}
    \vspace{0.2cm}
    \caption{Ratio $R=\frac{\max[z]}{\max[\tilde{z}]}$ vs $\eta=\Delta_0/\Omega_0$ and $x=\tau\Omega_0$, for (a) $z=\Omega_R(t)$ and (b) $z=\Delta(t)$. $R=1$ corresponds to the case in which the maximum amplitudes of the original functions from Eqs.~(4) and (5), are equal to those for the SATD-modified pulses from Eqs.~\eqref{eq:ModPulses}. }
    \label{fig:placeholder}
\end{figure}

%Notice that the function $g_z(t)$ is arbitrary and we can find $\mu(t)$ from equation (10b), which gives
%\begin{equation}
%    \mu(t)=\arctan{\left( \frac{\Dot{\theta}\cos{\varphi}-\Dot{\varphi}\sin{\theta}\sin{\varphi}}{g_z+\Omega+\Dot{\varphi}(1-\cos{\theta})}\right)}
%\end{equation}
%To get $g_x(t)$ we need to calculate the time derivative of $\mu$. For the case of an orange-slice path, where $\varphi$ is defined as a constant parameter, we get
%\begin{equation}
%    \Dot{\mu}(t)=\left[\frac{\Ddot{\theta}(g_z+\Omega)-\Dot{\theta}(\Dot{g}_z+\Dot{\Omega})}{(g_z+\Omega)^2+\Dot{\theta}^2\cos^2{\varphi}}\right]\cos{\varphi}
%\end{equation}

%\vspace{0.2cm}

\subsection{Evolution operator and resulting gate}
Since the hamiltonian \eqref{eq:HDS} defined in the DS basis is diagonal, we can rewrite it as $H_{DS}=E_{DS}(t) S_z$, where
\begin{equation} \label{eq:E_DS}
    E_{DS}(t) = \sqrt{(\Omega+g_z)^2 + \Dot{\theta}^2 \cos^2{\varphi}}
\end{equation}
Then, the evolution of an initial state $\ket{\psi(t_i)}$ is given by $\ket{\psi(t)}=U^{DS}(t,t_i)\ket{\psi(t_i)}$, being
\small
\begin{equation}
    U^{DS}(t,t_i) = U^\dagger_{ad}(t) V^\dagger(t) \exp{-i\int^t_{ti} E_{DS}(t') dt' S_z} V(t_i) U_{ad}(t_i)
\end{equation}
\normalsize
the evolution operator at time $t$. Considering the cyclic path depicted in Figure 1(b) for $H_0(t)$, $t_i=0$ and a total evolution time $T=4\tau$, the resulting quantum gate is constructed as a sequence of three segments forming the OS path. The initial state starts at a polar angle \( \chi\) and evolves to \( \pi \) with azimuthal angle \( \varphi_1 \), then returns from \( \pi \) to \( 0 \) with a phase \( \varphi_2 \), and finally goes back from \( 0 \) to \( \chi \) along \( \varphi_1 \). Each segment is associated with a fixed value of the parameter \( \varphi \), and the total dressed-state evolution operator at the final time $T$ is given by the ordered product 
\begin{equation}
    U^{DS}(T,0) = U^{DS}_{\varphi_1}(T, t_2) \cdot U^{DS}_{\varphi_2}(t_2, t_1) \cdot U^{DS}_{\varphi_1}(t_1, 0)
\end{equation}
where $\theta(t_1)=\pi$ and $\theta(t_2)=0$.
Finally, we get: 
\begin{equation} \label{eq:U_DS}
    U^{DS}(T,0)=\begin{pmatrix}
      \cos{\gamma_{t}} +i \cos{\chi}\sin{\gamma_{t}} & ie^{-i \varphi_1} \sin{\chi}\sin{\gamma_{t}} \\
 ie^{i \varphi_1} \sin{\chi}\sin{\gamma_{t}} & \cos{\gamma_{t}} - i \cos{\chi}\sin{\gamma_{t}}  \end{pmatrix}
\end{equation} 
where $\gamma_t =  \gamma_g + \gamma_d$ is the total phase accumulated during the evolution. The first term corresponds to the geometric phase $\gamma_g=\pi-(\varphi_2-\varphi_1)$, which arises from the trajectory traced by the quantum state in the projective Hilbert space. The second term is the dynamical phase $\gamma_d=I_{\varphi_1}-I_{\varphi_2}$ with $I_{\varphi_1}=\int_0^{t_1} E_{DS}(t) dt +\int_{t_2}^T E_{DS}(t) dt$ and $I_{\varphi_2}=\int_{t_1}^{t_2} E_{DS}(t) dt$, that depends on the instantaneous energy of the state and the duration of the evolution. Figure 1(c) provides a graphical representation of the phases accumulated in each part of the path described in the DS frame. The minus sign between the integrals arises from the fact that the trajectory on Bloch's sphere is traversed in opposite directions during the outer and intermediate segments. 
%\vspace{0.2cm}

By comparing the DS gate with the geometric gate in Eq.~\eqref{eq:Ugeom}, we can observe that both share similar forms, but differ in the phase components. The DS operator corresponds to a dynamical gate, as it includes a dynamical phase. This arises because the STA technique employed does not enforce the system to evolve strictly along the adiabatic states of $H_0(t)$; instead, it only ensures that the initial and final states coincide with these, so the evolution obtained in the original frame is not necessarily cyclic (see Appendix B). Then, to obtain a purely geometric gate, the phase $\gamma_d$ must be somehow eliminated.

\section{Geometric quantum gates}

In this section we propose a function $g_z(t)$ designed to effectively suppress the dynamical component of the phase $\gamma_t(t)$ and enable the realization of purely geometric single-qubit gates. Subsequently, we analyze the robustness of the gates $U_z(\gamma_g)$ and $U_x(\gamma_g)$ generated through this STA protocol in the presence of pulse imperfections and decoherence effects induced by the environment. 
\begin{figure*}[t!]
\centering
\includegraphics[width=0.8\textwidth]{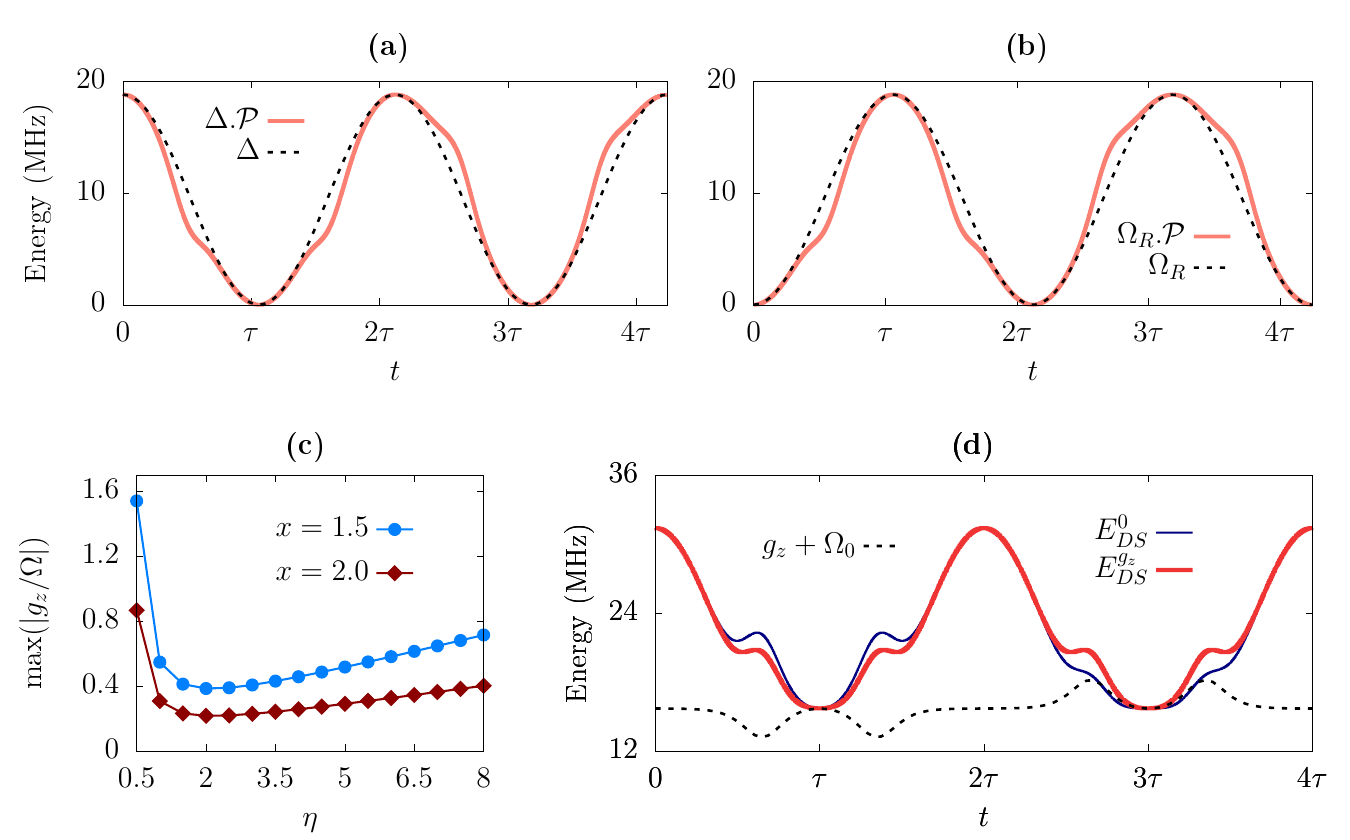}
\vspace{0.2cm}
\caption{Correction due to $g_z(t)$ from Eq.\eqref{eq:gz} for the original pulses (a) $\Delta(t)$ and (b) $\Omega_R(t)$ in terms of the rescaling factor $\mathcal{P}$ (see Eq.~\eqref{eq:P}); calculations correspond to the gate $U_z(\pi/2)$ with $\Delta_0=\Omega_0=2\pi \times 3 $ MHz and $x=2$. (c) Maximum value of the function $|g_z/\Omega|$ vs $\eta$ for two values of $x$; it can be seen that the function’s peak exhibits a minimum when $\eta=2$ for any $x$, and the highest values are reached when $\eta\ll1$. The plot also supports the analytical result shown in Eq.\eqref{eq:gz/Omega}, where a dependence on the evolution time decaying as $1/\tau^2$ is observed. (d) Energies $E_{DS}^0$ with $g_z = 0$, $E_{DS}^{g_z}$ and $g_z(t)+\Omega_0$; the comparison between the two DS energies shows that the role of $g_z$ is to symmetrize this function, ensuring that $I_{\varphi_1}=I_{\varphi_2}$ and, therefore, a vanishing dynamical phase $\gamma_d$. }
\label{fig:energy}
\end{figure*}

\subsection{Elimination of the dynamical phase}
As pointed out in the previous section, the system resulting from the DS protocol does not follow the adiabatic path during the whole evolution, acquiring a phase $\gamma_d(t)$ in the process. Therefore, in order to obtain a purely geometric gate, we must somehow impose the condition of a zero dynamical phase. One way to achieve this is by choosing $\varphi_1=-\varphi_2=\varphi_0$. However, this choice restricts the value of the geometric phase to $\gamma_g=\pi-2\varphi_0$ and does not allow the elimination of the exponential factors $e^{\pm i\varphi_1}$ in the non-diagonal terms of Eq.~\eqref{eq:U_DS}. As a result, the desired universal set of gates cannot be implemented using this method.
%\vspace{0.1cm}

Another way to solve this problem is to take advantage of the arbitrariness of the function $g_z(t)$, which can be used to enforce the condition $I_{\varphi_1} = I_{\varphi_2}$. 
%\textcolor{blue}{Here, it is important to notice that in the OS path construction the specific choice of $\theta(t)$ and $\Omega(t)$ ensures that the dynamics within each segment of the trajectory have the same functional form, and the instantaneous energy differs between segments only by a constant shift due to $\varphi$. This implies that the integral of the energy over any time interval (e.g., $[0,\tau]$, $[\tau,2\tau]$, etc.) yields the same result. Therefore,}. 
%Due to the pulse symmetry, the time dependence of the parameters $\Omega_R$ and $\Delta$ fulfill the relation $f_A(t)=f_B(t-2\tau)$, with $f$ being one the two parameters.
The shape of the original pulses $\Delta(t)$ and $\Omega_R(t)$, and consequently $\theta(t)$ and $\Omega(t)$, are symmetric with respect to $t=2\tau$ (midpoint of the evolution). As a result, the integral of any time-dependent function over the interval $[0,2\tau]$ yields the same result as the integral over $[2\tau,4\tau]$. Therefore, the condition to cancel the dynamical phase can be expressed as:
\begin{multline} \label{eq:EqualIntegrals}
\int_0^{2\tau} \sqrt{(\Omega + g_z^A)^2 + \cos^2 \varphi_1 \, \dot{\theta}^2} \, dt \\
= \int_0^{2\tau} \sqrt{(\Omega + g_z^B)^2 + \cos^2 \varphi_2 \, \dot{\theta}^2} \, dt
\end{multline}
where we consider the general case in which $g_z$ can take different forms for each $\varphi$-value. This equality can be rewritten as $\int^{2\tau}_0 [f_1(t)-f_2(t)]dt =0$, being $f_1$ and $f_2$ the square root on the left- and right-hand sides of Eq.~\eqref{eq:EqualIntegrals}, respectively. If this difference can be expressed as the time derivative of a function $F(t)$, the integral becomes
\begin{equation}
    \int_0^{2\tau} \frac{d}{dt}F(t)\ dt= F(2\tau)-F(0)=0
\label{eq:Int-dFdt}
\end{equation}
There exist many forms for $F(t)$ that satisfy this condition, each one leading to a family of functions $g_z$ that will depend on $\dot{\theta}$ and/or $\Omega$. 

In the following, we take $\varphi_1=0$ to obtain quantum gates analogous to Eq.~\eqref{eq:Ugeom} and, to ensure that these gates are purely geometric, we propose that Eq.~\eqref{eq:Int-dFdt} is satisfied for the trivial case where $F=0$ (i.e. $f_1=f_2$). For $g_z^B$ and $g_z^A$, one can consider any particular form, such as $g^A=0$ and $g_z^B \ne 0$, or $g^A\ne 0$ and $g_z^B = 0$. However, we have found that choosing $g_z^B=-g_z^A=g_z$ leads to a simpler solution:
\begin{equation} \label{eq:gz}
  g_z(t) = \alpha  \ \Dot{\theta}(t)^2/\ \Omega(t) 
\end{equation}
where $\alpha= \sin^2{(\varphi_2)} /4$ is a constant that depends only on the non-trivial value of the azimuthal angle.
%\vspace{0.1cm}

\textcolor{black}{
To gain physical insight into the role of \( g_z \), we recall that its effect on the control pulses manifests as a global scaling factor \( \mathcal{P} = \left(1 + \frac{g_z}{\Omega}\right) \) that modifies their profile. This factor depends on the speed of the evolution $\dot{\theta}$ according to:
}
\begin{eqnarray} \label{eq:P}
    \mathcal{P} = 1 + \alpha \left( \frac{\dot{\theta}}{\Omega} \right)^2
\end{eqnarray}
\textcolor{black}{
This dependence can be directly observed in Eq.~\eqref{eq:E_DS} and allows us to assess whether the correction remains close to unity or grows significantly larger, which is essential for determining the experimental feasibility of the protocol.} 

%\begin{equation}
%\int_{0}^{\tau} \sqrt{(\Omega+g_z^A)^2 + \cos^2\varphi_1 \ %\dot{\theta}^2} \ dt = \int_{0}^{\tau} \sqrt{(\Omega+g_z^B)^2 + %\cos^2\varphi_2\ \dot{\theta}^2} \ dt    
%\end{equation}

%Although there are several ways to define this function, we will impose that $g_z$ is such that cancels out the effect during the complete evolution time.
%We could also try to obtain a solution for $g_z$ in which not the integrand, but the integral on both sides are equated. In order to solve the problem, we can also propose a similar form of $g_z= \alpha_p \dot{\theta}/\bar{\Omega}$, with $\bar{\Omega}$ an average of the pulse energy $\Omega(t)$. Such expression is obtained starting with an expansion in order to compare the results of the integrals. We show in the Appendix both approximations and compared with numerical determination of $\alpha_P$. We found that the expression in Eq. \ref{eq:gz} is more simple, and also it fits better with the numerical calculations for short times, in which the adiabatic approximation cannot be used.    

In Figure~3 we analyze the effects of the SATD protocol on the system. Panels (a) and (b) illustrate how the scaling factor \( \mathcal{P} \) modifies the original time dependence of the driving parameters \( \Delta \) and \( \Omega_R \). The magnitude of this correction for each parameter depends on the chosen amplitudes $\Omega_0$ and $\Delta_0$, and the evolution time $\tau$. The plots correspond to $\Omega_0/2\pi = 3$ MHz, $\eta=1$ and $x=2$, and it can be seen that $\mathcal{P}$ neither modifies the maximum amplitude of the pulses nor produce abrupt changes in their profile. Now, to ensure this conditions are always met in the design of our gate, we must choose the system constants such that the ratio 
$g_z/\Omega$ remains small. This can be evaluated analytically by expressing this ratio as 
\begin{equation} \label{eq:gz/Omega}
    \frac{g_z}{\Omega}= \frac{4~\alpha~\pi^2}{x^2}~h(\eta,t)
\end{equation}
where $h(\eta,t)$ is defined by the time dependence of $\Delta$ and $\Omega_R$ within each time interval (see Eqs.~\eqref{eq:Delta(t)} and \eqref{eq:OmegaR(t)}). Equation~\eqref{eq:gz/Omega} shows that the maximum of the ratio decays as the inverse square of $x$, and hence of $\Omega_0$ and $\tau$. In Fig.~2(c), we show the peak value of $|g_z/\Omega|$ vs $\eta$ for two evolution times. The figure demonstrates that this peak increases for shorter $\tau$ (smaller $x$), as it was expected, and it reaches a minimum when $\eta=2$.
Finally, in Fig. 2(e) we compare the energies $E_{\mathrm{DS}}^0$ (blue thin line, corresponding to $g_z = 0$) and $E_{\mathrm{DS}}^{g_z}$ (red thick line) with $g_z$ given by Eq.~\eqref{eq:gz}, for the gate $U_z^{DS}(\pi/4)$. The function $g_z$ was also plotted (dashed lines), shifted by $\Omega_0$ along the vertical axis, to provide a clearer insight of its effect on the energy:
it is shown how $g_z$ symmetrizes $E_{DS}$ around $t=T/2 $, effectively canceling the difference between the energy integrals before and after the midpoint of the evolution.

%\textcolor{red}{The results discussed on this section, demonstrate that the corrections arising from $g_z$ induce pulse modifications that are of the same order of magnitude as the original pulses, and therefore are experimentally implementable. The parameters $\eta=\Delta_0/\Omega_0$ and $x=\tau \Omega_0$ define the amplitude of the re-scaling of the pulses. Also, since $g_z$ is proportional to $\dot{\theta}^2$, in the adiabatic limit such corrections tend to vanish, which is a necessary condition in every STA technique. Finally, we point out that it is clear that adding $g_z$ represents a re-scaling to the original pulses, aside from the required pulse corrections already included in the Dressed-state formalism, we can ensure a purely non-adiabatic geometric gate.}    

The results of this section have demonstrated that the addition of a non-trivial $g_z$ to the SATD protocol induces pulse modifications that are of the same order as the original ones, making them experimentally feasible and, in this case, providing the realization of a purely non-adiabatic geometric gate.
The parameters $\eta$ and $x$ set the amplitude of this re-scaling. Moreover, $g_z \propto \dot{\theta}^2$ so these corrections vanish in the adiabatic limit, as it is required in any STA protocol.

\subsection{Gate robustness under systematic errors and decoherence}

In this section, we examine how resilient our quantum gates are to different types of errors, which is a crucial requirement for the successful implementation of quantum algorithms. In the case of NV center-based systems, the main source of decoherence is the magnetic noise from the surrounding environment, which can cause fluctuations in the local magnetic field experienced by the NV electron spin. Furthermore, we must consider systematic errors arising from imperfections in the experimental setup, control pulses or calibration inaccuracies. 

To assess the effectiveness of the quantum gates proposed in this work, we compute their average fidelity defined as:
\begin{equation} \label{eq:F}
    F = \frac{|Tr(M)|^2 + Tr(M.M^\dagger)}{d (d+1)}
\end{equation}
where $d$ is the dimension of the system and $M = U_{0}^\dagger.U_{r}$, being $U_0$ the ideal geometric gate and $U_r$ the real one obtained via numerical calculations.
\vspace{0.2cm}

\begin{figure}[t!]
    \centering
    \includegraphics[width=1.0\linewidth]{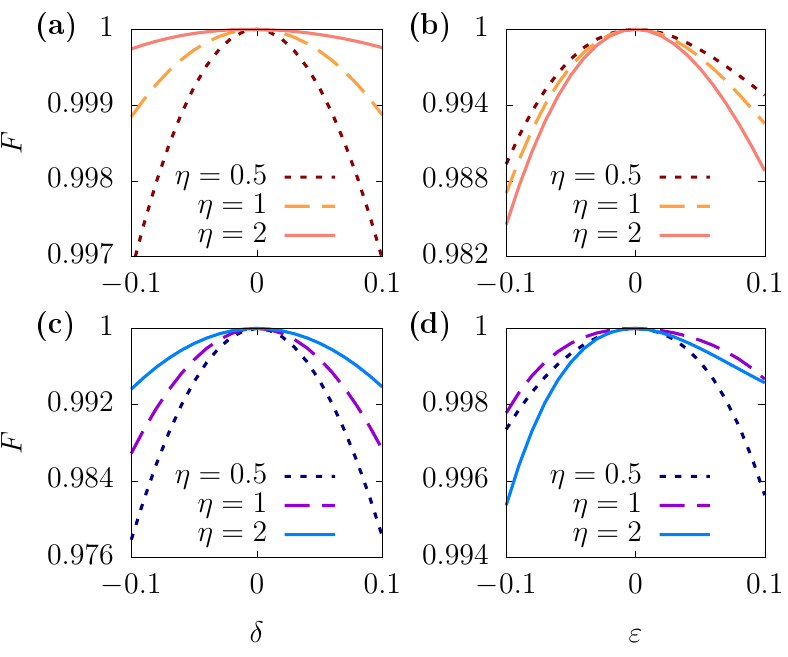}
    \caption{Gate fidelity vs systematic errors $\delta$ and $\varepsilon$, for different $\eta$-values. Panels (a) and (b) correspond to the NOT gate, and panels (c) and (d) to the S gate. Calculations were performed with parameters $\Omega_0/2\pi=3$ MHz and $\tau=2/\Omega_0$.}
    \label{fig:SE-delta-eps}
\end{figure}

We begin by defining the Hamiltonian $H_{SE}(t)$, which accounts for systematic errors as static deviations from the STA pulses $\Tilde{\Delta}$ and $\Tilde{\Omega}_R$. It is given by:
%\vspace{-0.2cm}
\small
\begin{equation}
   H_{SE}(t) =\frac{\hbar}{2}\begin{pmatrix}
        \Tilde{\Delta}(t)+\delta\Omega_0  & (1+\varepsilon)  \Tilde{\Omega}_R(t)e^{-i\Tilde{\varphi}(t)} \\
      (1+\varepsilon)  \Tilde{\Omega}_R(t)e^{i\Tilde{\varphi}(t)} & - \Tilde{\Delta}(t)-\delta\Omega_0
    \end{pmatrix}
\end{equation}
\normalsize
where $\delta$ is the qubit-frequency variation and $\varepsilon$ the driving deviation. In Figure 4 we present the gate fidelity \eqref{eq:F} as a function of the error parameters for $U_x(\pi/2)$ (NOT gate, panels a–b) and $U_z(\pi/2)$ (S gate, panels c–d). Each curve corresponds to a different value of $\eta$, which sets the amplitudes of the original pulses. Guided by the analysis in Figs. 2 and 3, we choose $\tau=2/\Omega_0$ with $\Omega_0/2\pi=3$ MHz. For $\eta = 2$, the decay of $F$ with detuning error $\delta$ is less pronounced than for the other values, in both gates. In contrast, for fidelity as a function of $\varepsilon$, this behavior is no longer observed, although the differences between $\eta$-values remain of the order of $10^{-3}$ even at the largest deviations.

From the results of Fig.~4 it is also clear that the two gates respond differently to systematic errors. For detuning errors $\delta$, the NOT-gate maintains significantly higher fidelities than the S-gate for all values of $\eta$. Conversely, when considering deviations in the Rabi frequency $\varepsilon$, the S-gate proves to be more robust, showing higher fidelities than the NOT operation in the same range of errors. These contrasting trends are consistent with the distinct physical mechanisms underlying each operation: the S-gate relies on the precise accumulation of a geometric phase, making it more sensitive to $\delta$, while the NOT-gate corresponds to a population inversion driven by the Rabi frequency, and is therefore more affected by errors in $\varepsilon$. The top panels of Figure 5 illustrate the combined effect of both errors for $\eta=2$, showing fidelities above 0.96 even for deviations as large as 15\% from the ideal amplitudes.

\begin{figure}[t!]
\centering
%\vspace{-0.6cm}
\includegraphics[width=1.\linewidth]{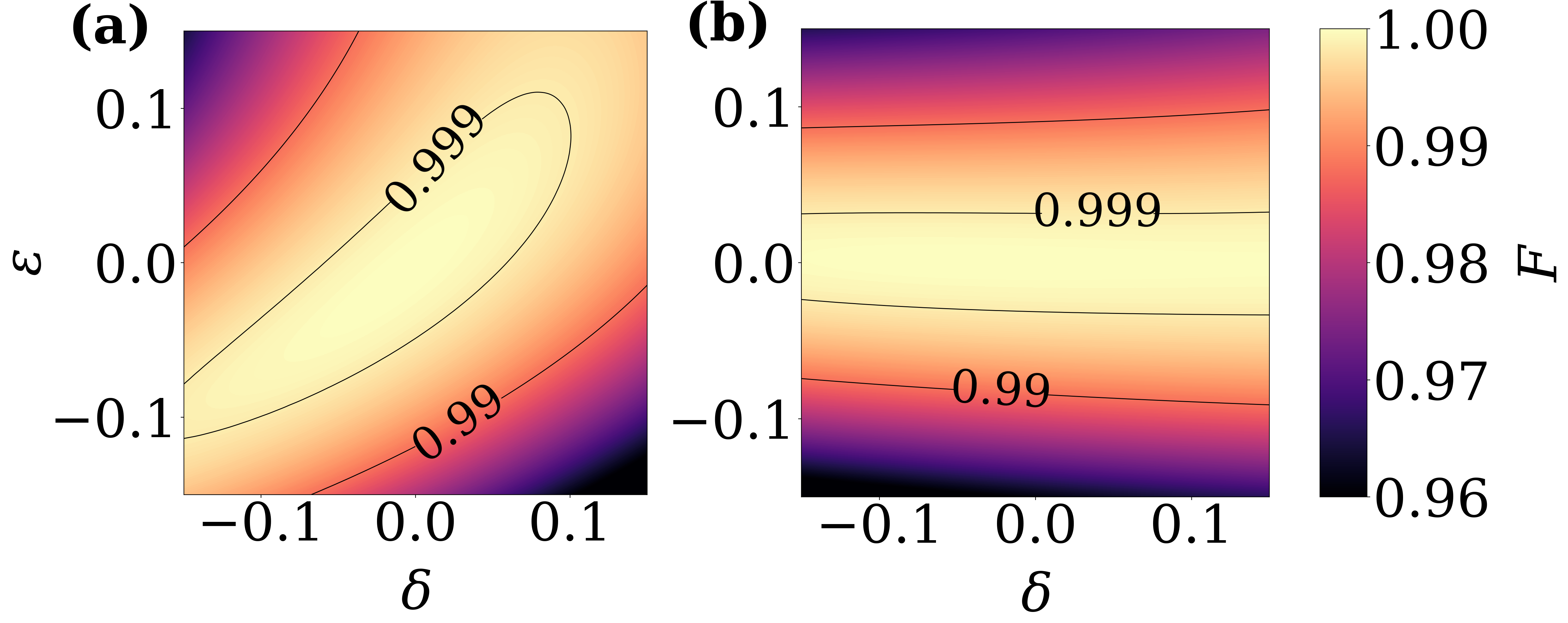}
\vspace{-0.4cm}

\includegraphics[width=1.\linewidth]{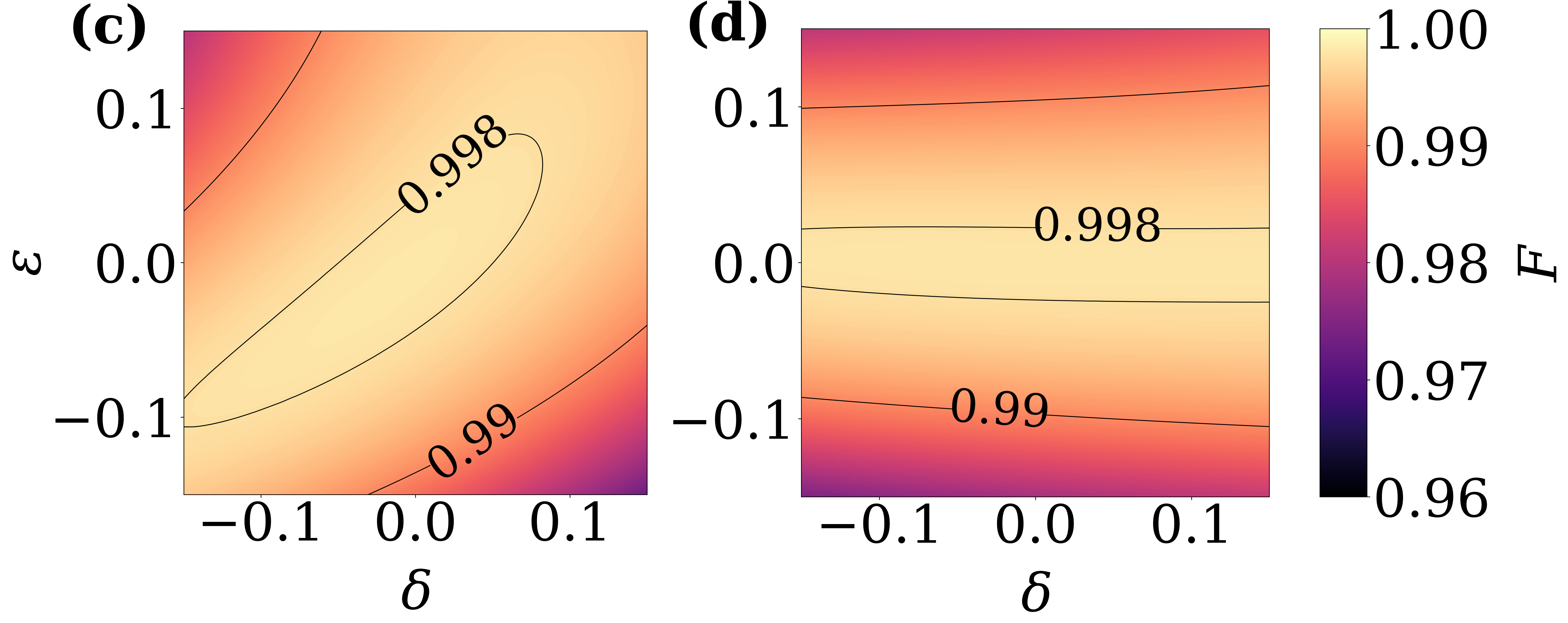}

\vspace{0.2cm}
\caption{Gate fidelity vs the qubit-frequency shift $\delta$ and driving deviation $\varepsilon$ for single-qubit gates (a) S and (b) NOT, and for two-qubit gates (c) CS and (d) CNOT. The parameters used are the same as in Fig.~4 and $A_{hf}=2\pi \times130$ MHz.}
\label{fig:SE}
\end{figure}

We also study the environmental decoherence on the $S-$gate, mainly due to interactions with nearby nuclear and electronic spins. Numerical simulations are performed using the Lindblad master equation \cite{Lindblad1976}, which has the form:
\begin{equation}
    \Dot{\rho} = - i[H_{SE}(t),\rho] + \kappa_1 \mathcal{M}(b_{-}) + \kappa_2  \mathcal{M}(b_{z})
\end{equation}
being $\rho$ the density matrix operator and $\kappa_1$ and $\kappa_2$ the decay and dephasing rates, respectively. The Lindblad operator is $\mathcal{M}(b)=2b \rho b^\dagger -  \lbrace b^\dagger b,\rho \rbrace$, with $b_- = \ket{0}\bra{1}$ and $b_z = \ket{1}\bra{1}$. To analyze the performance of the gate under these effects, we redefine the average fidelity as $F=(2\pi)^{-1}\int_0^{2\pi} \bra{\psi_{ideal}(\theta)}\rho_{real}\ket{\psi_{ideal}(\theta)}d\theta$, where $\rho_{real}$ is the density operator obtained from the numerical simulations, and $\ket{\psi_{ideal}(\theta)}$ is the final state resulting from the action of the ideal gate on the initial state $\ket{\psi_{0}(\theta)}=\cos\theta \ket{0}+\sin\theta \ket{1}$. Integration is carried out numerically over 1001 initial states, with $\theta$ values uniformly distributed in the interval $[0,2\pi]$. 

%For simplicity, we first analyze the case without systematic errors ($\delta = \varepsilon = 0$). The rates $\kappa_1$ and $\kappa_2$ are defined as $\kappa_1 = 1/T_1$ and $\kappa_2 = 1/T_\phi$, where $T_1$ is the longitudinal relaxation time and $T_\phi$ is the pure dephasing time. For the NV center at room temperature or below, typical values are $T_1 \sim 1{-}100$~ms and $T_\phi \sim 10{-}100~\mu$s \cite{Bar2013, Bradley2019}, with both times increasing significantly in isotopically enriched diamond. In Fig. 5 we analyze how these mechanisms affect the average gate fidelity, considering values of $\kappa_1$ and $\kappa_2$ associated with the previous time ranges. The infidelity $1-F$ is plotted as a function of $\kappa_1$ in panel (a) and as a function of $\kappa_2$ in panel (b). The dephasing-induced decoherence is at least two orders of magnitude stronger than the relaxation rate, leading to much more significant fidelity losses. In each plot, two values of $\tau = n/\Omega_0$ were considered, with $n=2$ and $3$. All curves display a linear growth of the infidelity with a smaller slope for the shorter evolution time; longer gate durations imply a longer exposure to environmental noise, which naturally enhances the effect of decoherence.

%For simplicity, we first analyze the case without systematic errors ($\delta = \varepsilon = 0$). 

The rates $\kappa_1$ and $\kappa_2$ are defined as $\kappa_1 = 1/T_1$ and $\kappa_2 = 1/T_\phi$, where $T_1$ is the longitudinal relaxation time and $T_\phi$ is the pure dephasing time. For the NV center at room temperature, typical values are $T_1 \sim 1{-}100$~ms and $T_\phi \sim 0.1{-}1$~ms \cite{Bar2013, Bradley2019, Stanwix2010}, with both times increasing significantly in isotopically enriched diamond \cite{Jahnke2012}. 
In Fig.~6 we analyze how these mechanisms affect the average gate fidelity, considering values of $\kappa_1$ and $\kappa_2$ associated with the previous time ranges. Since the relaxation rate associated with $\kappa_1$ is two orders of magnitude smaller than the dephasing rate $\kappa_2$, its impact on the fidelity loss is much less significant. Therefore, we fix an intermediate value $\kappa_1 = 5\times 10^{-4} \mu s^{-1}$ and plot the fidelity as a function of $\kappa_2$ for the S and NOT gates, in the presence of systematic errors with amplitudes $\delta = \epsilon = 0.05$. 
In the figure, a linear decay of the average fidelity of both gates can be observed, which is consistent with previous studies~\cite{NoncyclicGG(2021), Zhou-SATD}. Moreover, the slope of the curve is larger for the $S$ gate. This is because dephasing processes directly suppress the relative phase between the states $\ket{0}$ and $\ket{1}$, on which this gate relies. In contrast, the NOT gate is mainly affected by population relaxation, which occurs on longer timescales.

\begin{figure}[b!]
\centering
  \includegraphics[width=0.75\linewidth]{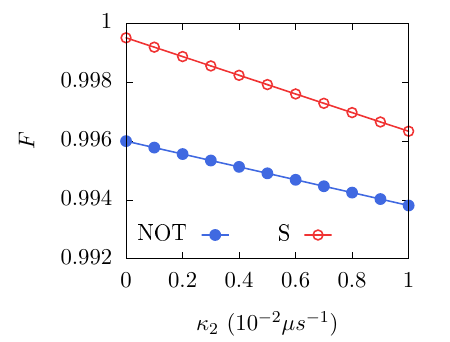}
  \vspace{0.1cm}
\caption{Gate Fidelity vs the decoherence rate $\kappa_2$ for the NOT (filled circles) and S (open circles) gates. We consider $\kappa_1=5\times10^{-4}$ $\mu s^{-1}$ and for $\eta$, $\Omega_0$ and $\tau$ we take the same values as in Figs.~4 and 5. We consider systematic errors with amplitudes $\delta=\varepsilon=0.05$. A linear decay of $F$ is observed in both cases, with a slope that is more pronounced for the S-gate. }
\label{fig:lindblad}
\end{figure}

We also studied the variation of the fidelity of both gates as a function of the parameter $\eta$, considering the same other values as in Fig.~6. We found that the changes on $F$ are almost negligible, although a slight improvement is observed as $\eta$ increases. In all cases, the fidelities remain above 0.99, which further confirms the robustness of geometric gates against different types of errors.

%In each plot, two values of $\tau = n/\Omega_0$ were considered, with $n=2$ and $3$. All curves display a linear growth of the  infidelity with a smaller slope for the shorter evolution time; longer gate durations imply a longer exposure to environmental noise, which naturally enhances the effect of decoherence.

%Finally, to analyze the combined effect of both decoherence rates and systematic errors on the average fidelity, we calculated $F$ for the NOT-gate with parameters \(\Omega_0 = 2\pi \times 3\,\mathrm{MHz}\), $\tau=2/\Omega_0$, \(\kappa_1 = 5 \times 10^{-4}\,\mu s^{-1}\), \(\kappa_2 = 5 \times 10^{-2}\,\mu s^{-1}\) and $\delta=\varepsilon=0.05$. Under these conditions, the resulting fidelity is 98.4\%, demonstrating that even in a strong fluctuations regime, the STA-based geometric gates maintain a high performance.

%\newpage
\subsection{Non-trivial two-qubit gates}

In this section we construct a nontrivial two-qubit gate by applying the STA protocol designed for a single-qubit to the system of an NV center strongly coupled to a nearby $^{13}C$ atom (spin $I=1/2$). The basis states in this case are of the form $\ket{m_s,m_I}$ with $m_s=0,-1$ for the electron spin and $m_I=\downarrow,\uparrow$ representing the nuclear spin. The full hamiltonian of the system in the secular approximation is given by \cite{Clerk-TQG, Finsterhölzl2025}
\begin{equation} \label{H_NV-C13}
    H_{NV-^{13}C}= D S_z^2+\gamma_eB_z~ S_z+\gamma_nB_z~I_z+A_{hf} ~S_zI_z
\end{equation}
being $\gamma_e$ and $\gamma_n$ the gyromagnetic ratio of the electron and nuclear spin, respectively, and $D$ is the zero-field energy gap.
We work in the interaction picture with respect to the nuclear Zeeman splitting by applying the unitary transformation \( U_n = e^{-i \gamma_n B_z I_z} \).

As in the single-qubit case, we drive the system with a MW field that perturbs only the electronic spin. In the rotating-wave approximation, the resulting Hamiltonian, written in the basis $\{\ket{0\downarrow},\ket{1\downarrow},\ket{0\uparrow},\ket{1\uparrow}\}$, takes the form  
\begin{equation} \label{eq:Htq}
\begin{split}
\small
    H_{tq}&=\frac{\hbar}{2} \begin{pmatrix}
        \Delta_{tq} && \Omega_R e^{-i\varphi} && 0 && 0 \\
        \Omega_Re^{i\varphi} && -\Delta_{tq} && 0 && 0 \\
        0 && 0 && \Delta_{tq} && \Omega_Re^{-i\varphi} \\
        0 && 0 && \Omega_R e^{i\varphi} && -\Delta_{tq} + 2 A_{hf} 
    \end{pmatrix},
\end{split}
\end{equation}
where $\Delta_{tq}$ and $\Omega_R$ are time-dependent functions modeled as in Eqs.~(4) and (5).  
Notice that this hamiltonian has a block-diagonal structure consisting of two uncoupled $2\times2$ sub-hamiltonians. The upper block, $h_1$, leads to an evolution operator identical to that previously obtained for a single qubit, reproducing the same dynamics. In contrast, the dynamics of the lower block, $h_2$, depend strongly on the value of the hyperfine coupling; if $A_{hf}\gg\Omega_R$, the evolution under $h_2$ becomes trivial, as no transitions occur between the states $\ket{0\uparrow}$ and $\ket{1\uparrow}$. This behavior is illustrated in Fig.~6, where we show how the gate infidelity $1-F$ diminishes as $A_{hf}$ increases. 
Consequently, the two-qubit gate implemented in this scheme can be written as  
\begin{equation} 
    U_{tq}(\chi,\gamma_g) = \ket{0}\bra{0} \otimes U_{sq}(\chi,\gamma_g) + \ket{1}\bra{1} \otimes I,
\end{equation}
where $U_{sq}$ is the single-qubit gate of Eq.~\eqref{eq:Ugeom} and $I$ is the $2\times2$ identity operator. This implementation realizes then a controlled gate, were $h_2$ acts as the control qubit.

\begin{figure}[t!]
    \centering
    \includegraphics[width=0.75\linewidth]{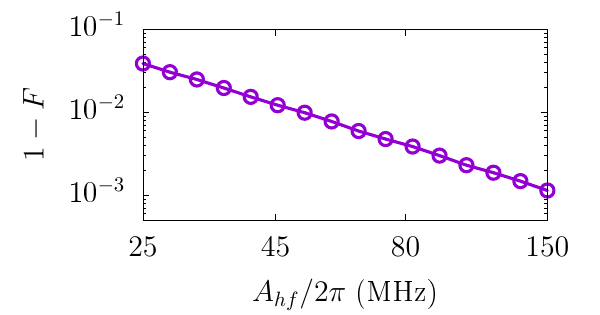}
    \vspace{0.2cm}
    \caption{Infidelity $1-F$ of the CS gate as a function of the hyperfine coupling $A_{hf}/2\pi$, that represents the interaction between the NV center and the nuclear spin of a nearby $^{13} C$ atom.}
    \label{fig:placeholder}
\end{figure}

The same STA protocol designed for a single-qubit gate can be implemented in this system. 
In Fig. 4(b) and (c) we show the performance of the CS and CNOT gates obtained with SATD under systematic errors. Here we take $A_{hf}/2\pi=130$ MHz, a value previously reported in the literature as experimentally feasible \cite{Felton2009,Rao2016}. The maximum fidelities found under small perturbations (less that 5\% deviations) are around 99.8\% in both cases. In both plots, the fidelity exhibits the same behavior under $\delta$ and $\varepsilon$ as in the single-qubit gates. However, in this case, the two-qubit gates retain higher fidelity values for larger deviations, but lower ones as we approach the ideal case with $\delta=\varepsilon=0$; this directly stems from the fixed value of the hyperfine coupling, which is constrained by experimental limitations.

\section{Conclusions}
%\textcolor{blue}{
%We proposed an efficient scheme for the realization of a universal set of fast geometric quantum gates by means of the shortcuts-to-adiabaticity SATD technique. In this approach, a suitable dressed-state basis is introduced together with a correction term that depends on two functions, $g_x(t)$ and $g_z(t)$, where the former is fixed by the STA conditions while the latter remains arbitrary. We showed that this protocol inevitably introduces an additional dynamical phase during the evolution. However, this unwanted contribution can be fully canceled through an appropriate choice of $g_z(t)$, thus enabling the implementation of purely geometric gates.   
%}

We proposed an efficient scheme for the realization of a universal set of fast geometric quantum gates by means of the shortcuts-to-adiabaticity SATD protocol. 
The strategy consists on adding a corrective Hamiltonian and performing a double unitary transformation on the system, where we define a convenient dressed-state basis where a trivial evolution is enforced. Within this framework, the system is able to follow the desired adiabatic dynamics in the real space within arbitrarily short times. The optimized pulses depend on a function 
$g_x(t)$, which couples the adiabatic states and governs the desired state evolution, and on another function $g_z(t)$ which does not directly affect the adiabatic dynamics but can be exploited to enhance the protocol, depending on the specific applications.
In this work, we have shown that the latter can be modeled to cancel the dynamical phases acquired during the evolution, thereby enabling the implementation of purely geometric gates.

Here, we focused on the particular case of a closed trajectory of the orange-slice type, which has been broadly studied and allowed us to illustrate the implementation and advantages of the proposed SATD protocol. However, it is important to emphasize that the protocol itself is general and can, in principle, be adapted to any path of interest. Research is currently underway to extend this approach to a broader class of trajectories, including open paths, in order to explore the full potential and flexibility of the this framework. Results from these ongoing studies will be reported in future work.

Numerical simulations demonstrated that the single-qubit gates designed with this method retain fidelities above 99.9\% under systematic errors with an intermediate amplitude, and over 99.4\% even with a strong environmental decoherence, confirming their experimental feasibility in NV centers and other solid-state platforms. Moreover, we extended the protocol to construct nontrivial two-qubit gates in a system consisting of an NV center coupled to a nearby nuclear spin via hyperfine interaction, showing that high-fidelity operations can also be achieved in this setting.  

Overall, our results demonstrate the feasibility of combining the intrinsic robustness of geometric quantum computation with the speed of shortcuts-to-adiabaticity techniques, paving the way toward scalable and high-fidelity quantum computing applications.  

%We have proposed an efficient scheme for the implementation of high-fidelity geometric quantum gates in two-level systems based on NV centers in diamond. Our approach relies on a STA mechanism based on dressed states, which enables robust and fast quantum operations. A key feature of the protocol is its dependence on an arbitrary function, offering a high degree of flexibility to tailor the control fields according to specific application requirements. 

\section*{Acknowledgement}
M.E.R, A.F and S.S.G. acknowledge CONICET 
(PIP11220200100170) and partial financial support from
 SGCyT-UNNE (PI 20T001).

\appendix
\setcounter{figure}{0} % reinicia contador de figuras
\renewcommand{\thefigure}{A\arabic{figure}} 

\section{Effects of the MW Phase Smoothing on the fidelity of the SATD protocol}
In this appendix, we study the behavior of the STA protocol when considering a time-dependent phase of the form:
\begin{equation} \label{eq:ModPhi}
    \varphi(t, \sigma)= \frac{\varphi_2}{2} \left[1+\tanh{\left(\frac{t-2\tau}{\sigma}\right)}\right]
\end{equation}
where $\sigma$ corresponds to the characteristic time of the phase transition. Experimentally, it is limited by the bandwidth of the microwave control electronics; for typical Rabi amplitudes of 10–20~MHz, $\sigma$ values of 5–10~ns are easily achievable~\cite{Herb-2025,Han2025}.

\begin{figure}[h!]
    \centering
    \includegraphics[width=1.\linewidth]{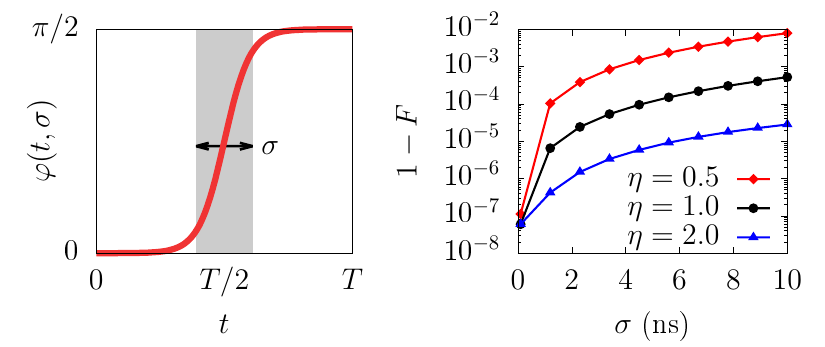}
    \caption{Left: Shape of the modulation \eqref{eq:ModPhi} of the MW phase for experimental implementation. Right: Average infidelity vs $\sigma $ for the S-gate resulting from the SATD strategy, for different $\eta$.}
    \label{fig:A1}
\end{figure}

In Fig.~A1, the shape of this modulation is shown for $\sigma = 10$~ns (left), along with the average infidelity $1-F$ of the $S$-gate obtained with the SATD protocol as a function of $\sigma$ (right). We consider different values of the detuning and Rabi amplitude, given by parameter $\eta$, where we set $\Omega_0/2\pi=3$ MHz. The second plot indicates that this modulation does not induce a significant decay in the gate fidelity for $\eta \ge 1$. Furthermore, this phase description ensures the continuity of the modified STA pulses throughout the evolution.

Finally, it is worth mentioning that for the parameter ranges where this modification has a significant impact on the gate fidelity, the protocol can be adjusted by considering a non-constant phase. Under these conditions, the new functions $g_x$ and $g_z$ take the form:
\begin{subequations}
    \begin{equation}
        g_x=-\dot{\mu} +\dot{\theta}\sin{\varphi} +\dot{\varphi} \sin\theta \cos\varphi
    \end{equation}
    \vspace{-0.9cm}
    \begin{equation}
        g_z=-\Omega+ -\dot{\varphi}(1-\cos\theta)+ \frac{1}{\tan{\mu}}(\dot{\theta}\cos{\varphi}-\dot{\varphi}\sin\theta \cos\varphi)
    \end{equation}
\label{eq:gxgz-phi(t)}
\end{subequations}
Although in this particular case the terms involving $\dot{\varphi}$ contribute only around $t = T/2$, these expressions demonstrate that more complex trajectories on the Bloch sphere lead to correction functions that strongly depend on the specific choice of both $\theta(t)$ and $\varphi(t)$.

\section{Modified pulses and Origin of the Dynamical Phase in real Space}

In Sections II and III of the main text, we discussed the emergence of the dynamical phase in the protocol and how to eliminate it; however, we did not explicitly address its direct relation to the SATD Hamiltonian, as defined in the original frame.  Therefore, in Fig. A2 we present a comparison between the corrected pulses, which include the contributions from both $g_x(t)$ and $g_z(t)$, and the original ones. Here, a new trajectory on the Bloch sphere can be defined, that is determined by a new polar angle
\begin{equation}
    \tilde{\theta}=\arctan{\left( \tilde{\Omega}_R/\tilde{\Delta}\right)}
\end{equation}
and a new azimuthal angle $\tilde{\varphi}$ from Eqs.~\eqref{eq:ModPulses}. In the lower panel, it can be seen that the initial amplitude of the new Rabi frequency is nonzero, which implies that the initial state is not exactly in $\ket{0}$. However, the evolution ends in the north pole of the sphere; this corresponds to an open trajectory that gives rise to the dynamical phase $\gamma_d$.

\begin{figure}[h!]
    \centering
    \includegraphics[width=0.85\linewidth]{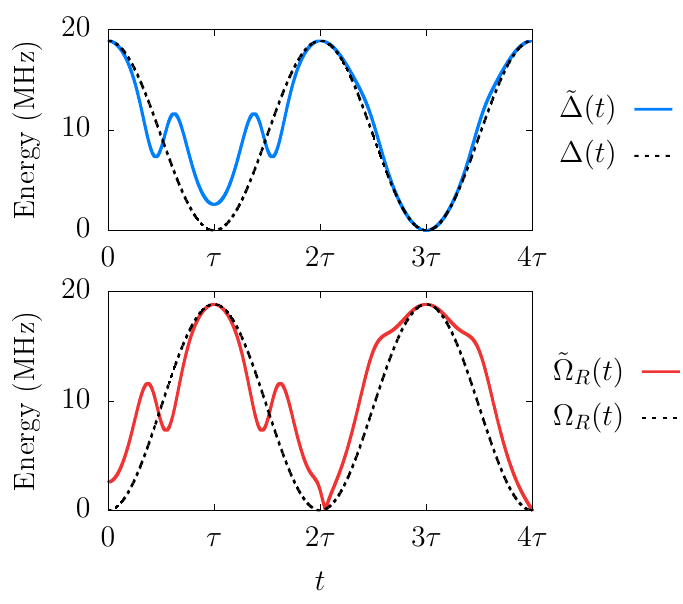}
    \caption{Comparison between the original (dashed lines) and SATD-modified (solid lines) pulses for the S-gate. The parameters $\Omega_0$, $\eta$ and $\tau$ are the same from Figs.~3(a) and (b) of the main text. The lower panel shows a modified pulse for $\Omega_R(t)$ that has a nonzero value at the initial time, which implies an open trajectory on the Bloch sphere and, consequently, the appearance of a dynamical phase accumulated during the evolution. }
    \label{fig:A2}
\end{figure}

%\newpage


\begin{thebibliography}{99}
% Papers GeomGates
\bibitem{Nielsen} M. A. Neilsen and I. L. Chuang, Quantum Computation and Quantum Information (Cambridge University Press, 2000).
\bibitem{Berry1984} M. V. Berry. Quantal phase factors accompanying adiabatic changes. Proc. R. Soc. A 392, 45–57 (1984).
\bibitem{Zanardi1999} Zanardi, P. and Rasetti, M. Holonomic quantum computation. Phys. Lett. A 264,
94–99 (1999).
\bibitem{Ekert(2000)} A. Ekert, M. Ericsson, P. Hayden, H. Inamori, J. A. Jones, D. K. L. Oi, and V. Vedral, Geometric Quantum Computation, J. Mod. Opt. 47, 2501 (2000).
\bibitem{GQG-Review2023} J. Zhang, T. Ha Kyaw, S. Filipp, L. Kwek, E. Sjöqvist and Di. Tong. Geometric and Holonomic Quantum Computation, Physics Reports 1027 (2023).
\bibitem{Zhu2002} S. L. Zhu and Z. D Wang. Implementation of universal quantum gates based on nonadiabatic geometric phases. Phys. Rev. Lett. 89, 097902 (2002).
\bibitem{Wu2013} H. Wu, E. M. Gauger, R. E. George, M. Möttönen, H. Riemann, N. V. Abrosimov, P. Becker, H.-J. Pohl, K. M. Itoh, M. L. W. Thewalt, and J. J. L. Morton, Geometric phase gates with adiabatic control in electron spin resonance, Phys. Rev. A 87, 032326 (2013).

\bibitem{Lupo2007} Lupo, C., Aniello, P., Napolitano, M. and Florio, G. Robustness against parametric noise of nonideal holonomic gates. Physical Review A, 76(1), 012309 (2007).

\bibitem{Chen2020} Chen, T. and Xue, Z.-Y. High-fidelity and robust geometric quantum gates that outperform dynamical ones. Physical Review Applied, 14(6), 064009 (2020).



%Papers STA y DS
\bibitem{STAreview(2019)} D. Guéry-Odelin, A. Ruschhaupt, A. Kiely, E. Torrontegui, S. Martínez-Garaot, and J. G. Muga, Shortcuts to adiabaticity: Concepts, methods, and applications. Rev. Mod. Phys. 91, 045001 (2019).
\bibitem{BerryTQD} Berry M. V. Transitioless quantum driving. J. Phys. A, 42 (2009) 365303.
\bibitem{LR-Invariants(2012)} A. Ruschhaupt, Xi Chen, D. Alonso and J. G. Muga. Optimally robust shortcuts to population inversion in two-level quantum systems. New J. Phys. 14 093040 (2012).
%\bibitem{}

\bibitem{PRA93(2016)} Zhen-Tao Liang, Xianxian Yue, Qingxian Lv, Yan-Xiong Du, Wei Huang, Hui Yan, and Shi-Liang Zhu. Proposal for implementing universal superadiabatic geometric quantum gates in nitrogen-vacancy
centers. Phys. Rev. A 93, 040305(R) (2016).
\bibitem{NoncyclicGG(2021)} Li-Na Ji, Cheng-Yun Ding, Tao Chen, and Zheng-Yuan Xue. Noncyclic Geometric Quantum Gates with Smooth Paths via Invariant-based Shortcuts. Adv. Quantum Technol., 4: 2100019 (2021).
\bibitem{Lv2020} Lv, Q.-X., Liang, Z.-T., Liu, H.-Z., Liang, J.-H., Liao, K.-Y. and Du, Y.-X. Noncyclic geometric quantum computation with shortcut to adiabaticity. Physical Review A, 101(2), 022330 (2020).


%STIRAP con SA-TD
\bibitem{Baksic(2016)} A. Baksic, H. Ribeiro, and A. A. Clerk.
Speeding up Adiabatic Quantum State Transfer by Using Dressed States,
Phys. Rev. Lett. 116, 230503 (2016).
\bibitem{Ribeiro-Clerk(2019)} Ribeiro, H. and Clerk, A. A. Accelerated adiabatic quantum gates: Optimizing speed versus robustness. Physical Review A, 100(3), 032323 (2019).
\bibitem{Coto-SATD} R. Coto, V. Jacques, G. Hétet, and J. R. Maze. Stimulated Raman adiabatic control of a nuclear spin in diamond. Phys. Rev. B 96, 085420 (2017).
\bibitem{Zhou-SATD} Xiao Zhou, Bao-Jie Liu, L-B Shao, Xin-Ding Zhang and Zheng-Yuan Xue. Quantum state conversion in opto-electro-mechanical systems via shortcut to adiabaticity. Laser Phys. Lett. 14, 095202 (2017).
\bibitem{Liu(2017)} Bao-Jie Liu, Zhen-Hua Huang, Zheng-Yuan Xue, and Xin-Ding Zhang. Superadiabatic holonomic quantum computation in cavity QED. Physical Review A, vol. 95 (2017).
\bibitem{Zhang2020} J. Zhang, S. S. Hegde, and D. Suter, 
Efficient implementation of a quantum algorithm in a single nitrogen-vacancy center of diamond, 
Phys. Rev. Lett. 125, 030501 (2020).




% GQG in Supercondutors
\bibitem{Li2021} S. Li, J. Xue, T. Chen, Z. Xue. High-Fidelity Geometric Quantum Gates with Short Paths on Superconducting Circuits. Advanced Quantum Technologies, 4(3), 2000140 (2021).
\bibitem{Yang2023} Yang, X.-X., Guo, L.-L., Zhang, H.-F., Du, L., Zhang, C., Tao, H.-R., Chen, Y., Duan, P., Jia, Z.-L., Kong, W.-C., and Guo, G.-P. Experimental implementation of short-path nonadiabatic geometric gates in a superconducting circuit. Physical Review Applied, 19(4), 044076 (2023).
\bibitem{Wang2018} T. Wang, Z. Zhang, L. Xiang, Z. Jia, P. Duan, W. Cai, Z. Gong, Z. Zong, M. Wu, J. Wu, L. Sun, Y. Yin, and G. Guo. The experimental realization of high-fidelity ‘shortcut-to-adiabaticity’ quantum gates in a superconducting Xmon qubit,  New J. Phys. 20, 065003 (2018).
\bibitem{Liang2024} Liang, Y., and Xue, Z.-Y. Nonadiabatic geometric quantum gates with on-demand trajectories. Physical Review Applied, 21(6), 064048 (2024).



% GQG in quantum dots
\bibitem{Ma2024} Ma, R.-L., Li, A.-R., Wang, C., Kong, Z.-Z., Liao, W.-Z., Ni, M., Zhu, S.-K., Chu, N., Zhang, C., Liu, D., Cao, G., Wang, G.-L., Li, H.-O., and Guo, G.-P. Single-spin-qubit geometric gate in a silicon quantum dot. Physical Review Applied, 21(1), 014044 (2024). 
% GQG in Rydberg atoms
\bibitem{Zhao2017} P. Z. Zhao, X.-D. Cui, G. F. Xu, E. Sjoqvist, and D. M. Tong. Rydberg-atom-based scheme of nonadiabatic geometric quantum computation. Phys. Rev. A 96, 052316 (2017).
\bibitem{Jin2024} Jin, Z.-Y., and Jing, J. Geometric quantum gates via dark paths in Rydberg atoms. Phys. Rev. A, 109(1), 012619 (2024).


% GQG in NV centers
\bibitem{Cheng2021} Cheng, J.-J. and Zhang, L. Implementing conventional and unconventional nonadiabatic geometric quantum gates via SU(2) transformations. Physical Review A, 103(3), 032616 (2021).
\bibitem{Chen2025-arxiv} Chen, S.-Q., Duan, Q.-T., Zhang, C. and Lu, H. Multiple-noise-resilient nonadiabatic geometric quantum control of solid-state spins in diamond (2025). arXiv preprint arXiv:2508.12221.


\bibitem{Doherty2013} Doherty, M. W., Manson, N. B., Delaney, P., Jelezko, F., Wrachtrup, J., and Hollenberg, L. C. L. (2013). The nitrogen-vacancy colour centre in diamond. Physics Reports, 528(1) (2013).
% Experiments
\bibitem{Arroyo2014} Arroyo-Camejo, S., Lazariev, A., Hell, S. et al. Room temperature high-fidelity holonomic single-qubit gate on a solid-state spin. Nat Commun 5, 4870 (2014).
\bibitem{Sekiguchi2017} Sekiguchi, Y., Niikura, N., Kuroiwa, R. et al. Optical holonomic single quantum gates with a geometric spin under a zero field. Nature Photon 11, 309–314 (2017).
\bibitem{Bartling2025} Bartling, H. P., Yun, J., Schymik, K. N., van Riggelen, M., Enthoven, L. A., van Ommen, H. B., Babaie, M., Sebastiano, F., Markham, M., Twitchen, D. J. and Taminiau, T. H. Universal high-fidelity quantum gates for spin qubits in diamond. Physical Review Applied, 23(3), 034052 (2025).
\bibitem{Kleißler2018} Kleißler, F., Lazariev, A. and Arroyo-Camejo, S. Universal, high-fidelity quantum gates based on superadiabatic, geometric phases on a solid-state spin-qubit at room temperature. Quantum Inf 4, 49 (2018).
\bibitem{Ma2023} Ma, Z., Xu, J., Chen, T., Zhang, Y., Zheng, W., Li, S., Lan, D., Xue, Z.-Y., Tan, X., and Yu, Y. Noncyclic nonadiabatic geometric quantum gates in a superconducting circuit. Physical Review Applied, 20(5), 054047 (2023).
\bibitem{Zu2014} Zu, C., Wang, WB., He, L. et al. Experimental realization of universal geometric quantum gates with solid-state spins. Nature 514, 72–75 (2014).


\bibitem{Lindblad1976} Lindblad, G. On the generators of quantum dynamical semigroups. Commun.Math. Phys. 48, 119–130 (1976).


\bibitem{Bar2013} Bar-Gill, N., Pham, L. M., Jarmola, A., Budker, D. and Walsworth, R. L. Solid-state electronic spin coherence time approaching one second. Nature Communications, 4(1), 1743 (2013).
\bibitem{Bradley2019} Bradley, C. E., Randall, J., Abobeih, M. H., Berrevoets, R. C., Degen, M. J., Bakker, M. A., Markham, M., Twitchen, D. J. and Taminiau, T. H. A ten-qubit solid-state spin register with quantum memory up to one minute. Physical Review X, 9(3), 031045 (2019).

\bibitem{Stanwix2010} Stanwix, P. L., Pham, L. M., Maze, J. R., Le Sage, D., Yeung, T. K., Cappellaro, P., Hemmer, P. R., Yacoby, A., Lukin, M. D.,and Walsworth, R. L. Coherence of nitrogen-vacancy electronic spin ensembles in diamond. Physical Review B, 82(20), 201201 (2010).
\bibitem{Jahnke2012} K. D. Jahnke, B. Naydenov, T. Teraji, S. Koizumi, T. Umeda, J. Isoya, F. Jelezko; Long coherence time of spin qubits in 12C enriched polycrystalline chemical vapor deposition diamond. Appl. Phys. Lett. 2; 101 (1) 012405 (2012).


% TQ gates 
\bibitem{Clerk-TQG} Setiawan, F., Groszkowski, P. and Clerk, A. A. Fast and robust geometric two-qubit gates for superconducting qubits and beyond. Physical Review Applied, 19(3), 034071 (2023).
\bibitem{Finsterhölzl2025} Finsterhölzl, R., Hannes, W.-R. and Burkard, G. High-Fidelity Entangling Gates for Electron and Nuclear Spin Qubits in Diamond. Physical Review B, 111(21), 214104 (2025).
\bibitem{Felton2009} Felton, S., Edmonds, A. M., Newton, M. E., Martineau, P. M., Fisher, D., Twitchen, D. J. and Baker, J. M. Hyperfine interaction in the ground state of the negatively charged nitrogen vacancy center in diamond. Physical Review B, 79(7), 075203 (2009).
\bibitem{Rao2016} Rao, K. R. K. and Suter, D. Characterization of hyperfine interaction between an NV electron spin and a first-shell $^{13}C$ nuclear spin in diamond. Physical Review B, 94(6), 060101 (2016).




% MM bandwidth
\bibitem{Herb-2025} Herb, K., Völker, L.A., Abendroth, J.M. et al. Quantum magnetometry of transient signals with a time resolution of 1.1 nanoseconds. Nat Commun 16, 822 (2025).
\bibitem{Han2025} Han, S., Ye, X., Zhou, X., Liu, Z., Guo, Y., Wang, M., Ji, W., Wang, Y. and Du, J. Solid-state spin coherence time approaching the physical limit. Science Advances, 11(9) (2025).

\end{thebibliography}
\end{document}